\documentclass[fleqn,twoside]{article}

\usepackage{amsmath,amssymb,verbatim}
\usepackage{graphicx}
\usepackage{pifont}
\usepackage{espcrc2}
\usepackage{cite}

\newcommand{\E}{\mathbf{E}}
\newcommand{\B}{\mathbf{B}}

\newcommand{\bra}[1]{\langle\,#1\,|}          % Dirac bra
\newcommand{\ket}[1]{|\,#1\,\rangle} 
\newcommand{\x}{\boldsymbol{x}}
\newcommand{\y}{\boldsymbol{y}}

\newcommand{\ud}{\mathrm{d}}
\newcommand{\be}{\begin{equation}}
\newcommand{\ee}{\end{equation}}
\newcommand{\bestar}{\begin{equation*}}
\newcommand{\eestar}{\end{equation*}}
\newcommand{\bi}{\begin{itemize}}
\newcommand{\ei}{\end{itemize}}
\title{Noncommutativity and the lightfront}

\author{Thomas Heinzl\address{School of Computing and Mathematics, University of Plymouth, PL4 8AA, UK.\\
\texttt{theinzl@plymouth.ac.uk}},
        Anton Ilderton\address{Department of Physics, Ume\aa\ University, SE--901 87 Ume\aa, Sweden.\\
\texttt{anton.ilderton@physics.umu.se}}\thanks{Presented by A.~I. at `LightCone 2009', ITA, S\~ao Jos\'e dos Campos, Brazil, 08--13/07/2009.},
        Mattias Marklund\address{Department of Physics, Ume\aa\ University, SE--901 87 Ume\aa, Sweden.\\
\texttt{mattias.marklund@physics.umu.se}}
      }
      
\begin{document}

\begin{abstract}
We discuss various limits which transform configuration space into phase space, with emphasis on those related to lightfront field theory, and show that they are unified by spectral flow. Examples include quantising in `almost lightfront' coordinates and the appearance of lightlike noncommutativity from a strong background laser field. We compare this with the limit of a strong magnetic field, and investigate the role played by lightfront zero modes.
\vspace{1pc}
\end{abstract}

\maketitle

\section{Introduction}
Consider the following systems and limits, which initially appear to be unrelated. First, the quantum mechanics of a particle in a magnetic field, in the limit that the magnetic field strength is very large. Second, field theory in coordinates which interpolate between quantisation surfaces of timelike and, in the limit, lightlike orientation. Although they appear very different, we will see that these two limits, and many others, can be given a unified description in terms of `spectral flow' -- this is a decoupling of states which transforms configuration space into phase space, and, as a result, introduces a noncommutativity into the theory \cite{Heinzl:2007ca}. Classically, this can be signalled by the limit in question transforming the relevant action from quadratic to linear in time derivatives, implying a change in canonical structure.

We will give several examples of spectral flow, beginning with the theory of a particle in a strong magnetic field, which now serves as a standard introduction to noncommutative theories \cite{Szabo:2001kg, Barbon:2001bi, Szabo:2004ic}. We will then see that exactly the same spectral flow mechanism is at work when we use interpolating, or `almost lightfront' coordinates to investigate lightfront field theory. After briefly mentioning two related limits in string and field theory, we discuss the appearance of lightlike noncommutativity within the theory of a particle in a strong {\it crossed} electromagnetic field. This is a low frequency approximation to the fields of an intense laser. We will see that despite the classical similarity between this limit and that of a strong magnetic field, the quantum limits are very different. In particular, a key role is played by lightfront zero modes in the strong crossed field limit.
\section{A particle in a magnetic field}
Our first example is a non--relativistic particle in a constant magnetic field of magnitude $B$, aligned in the $x^3$ direction \cite{landau:1930}. We give only the essential details as this topic is well covered in the literature \cite{Szabo:2001kg, Barbon:2001bi, Szabo:2004ic}. The action is
\be\label{lag1}
	S = \int\!\ud t\ \frac{m}{2}\dot\x^2 + e B \dot{x}^1 x^2 - V(\x)\;,
\ee
where $V$ is some potential describing other interactions. We consider the limit of a strong magnetic field, $eB\gg m^2$, such that we can neglect the kinetic term of the action; this limit may alternatively be thought of as taking $m\to0$. In this limit we are left with an action which is linear, rather than quadratic, in time derivatives,
\be
	S \to \int\!\ud t\ e B \dot{x}^1 x^2 - V(\x)\;.
\ee
Quantising this action, we would infer the equal time commutator and Hamiltonian,
\be\label{NCt}
	[x^1,x^2]= \frac{i}{eB} \quad\text{and}\quad H = V\;,
\ee
respectively. We have found {\it space--space} noncommutativity: the coordinates of the $(x^1,x^2)$ plane no longer commute. (The same conclusion is reached through a more careful treatment of the limit using Dirac brackets -- see \cite{Dunne:1989hv, Dunne:1992ew, Guralnik:2001ax, Jackiw:2001dj, Jackiw:2002wd, Horvathy:2002wc
} for more detail on this, and other, aspects of the limit.) Calculations are performed in this theory by making the replacement
\be
	x^2 \to -\frac{i}{eB}\frac{\partial}{\partial x^1}\;,
\ee
in $V(\x)$, which is just the `Peierls substitution' \cite{Peierls:1933}. To better understand the limit, we turn to the energy spectrum of the particle. Returning to (\ref{lag1}), and for simplicity setting $V=0$, the Hamiltonian is essentially that of a harmonic oscillator, frequency $eB/m$, with spectrum
\be\label{Enll}
	E(n) = \frac{eB}{m}\bigg(n+\frac{1}{2}\bigg) + \frac{p_3^2}{2m}\;.
\ee
Killing the free motion in the $x^3$--direction, so that the particle is confined to the $(x^1,x^2)$ plane, these $E(n)$ are the `Landau levels', which are infinitely degenerate with respect to $p_1$, a momentum component in the plane. Now, in our limit, we see from (\ref{Enll}) that the spectral gap $eB/m$ between the Landau levels increases. This leads to the quantum Hall effect: as the cost of accessing excited states increases, we find higher occupation numbers in the low--lying Landau levels. Thus, as $B$ increases (or $m$ decreases), excited states become inaccessible, and decouple from the theory during this spectral flow. As $B\to\infty$ only the ground state, $n=0$, is available. The limit therefore projects onto the lowest Landau level, and it is here that the particle is governed by the noncommutative theory (\ref{NCt}). We summarise these results by saying we have found `noncommutativity from spectral flow' \cite{Heinzl:2007ca}.
\section{Toward the lightfront}
\begin{figure*}[th!]
	\includegraphics[width=0.4\textwidth]{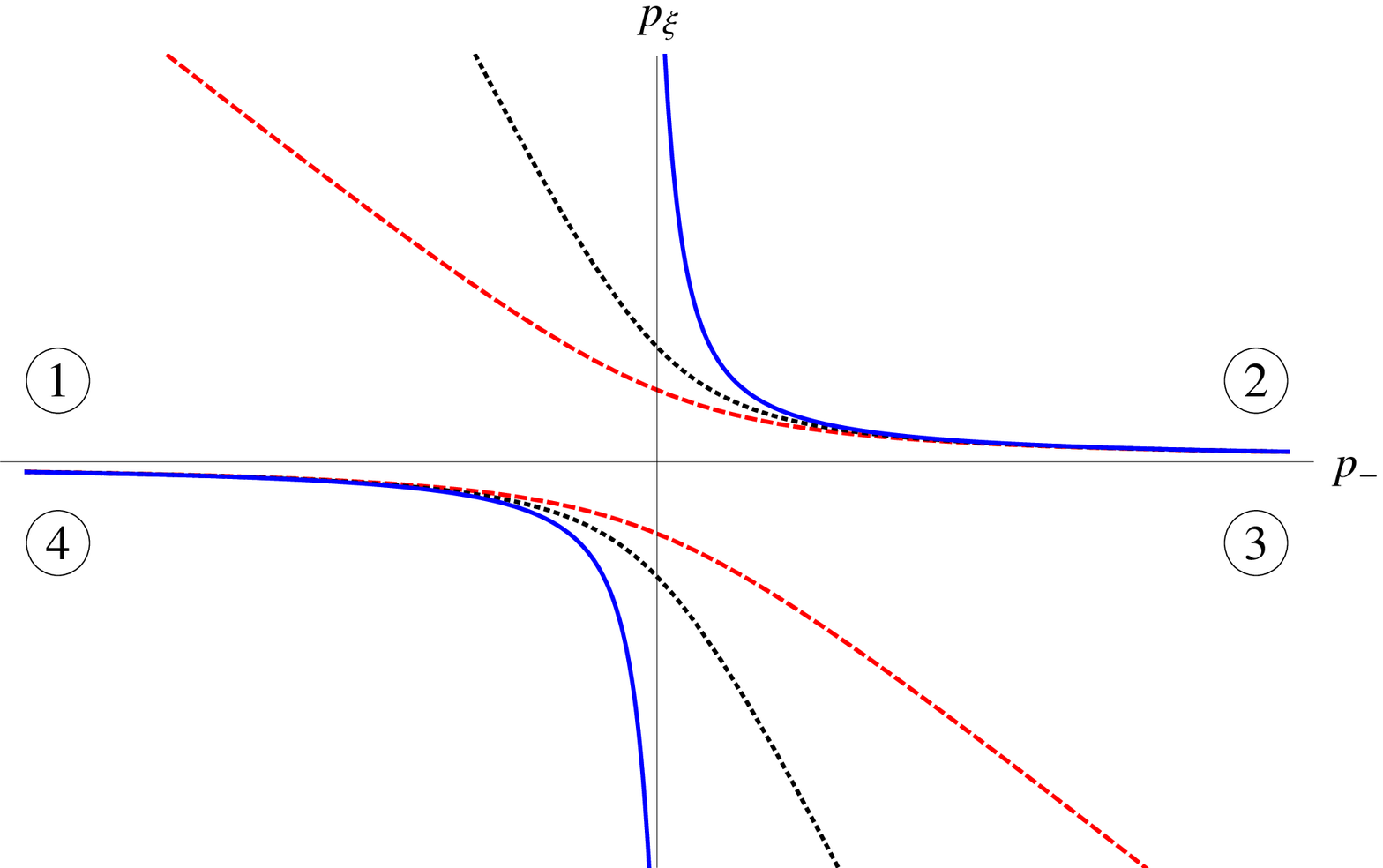}\hfill\includegraphics[width=0.4\textwidth]{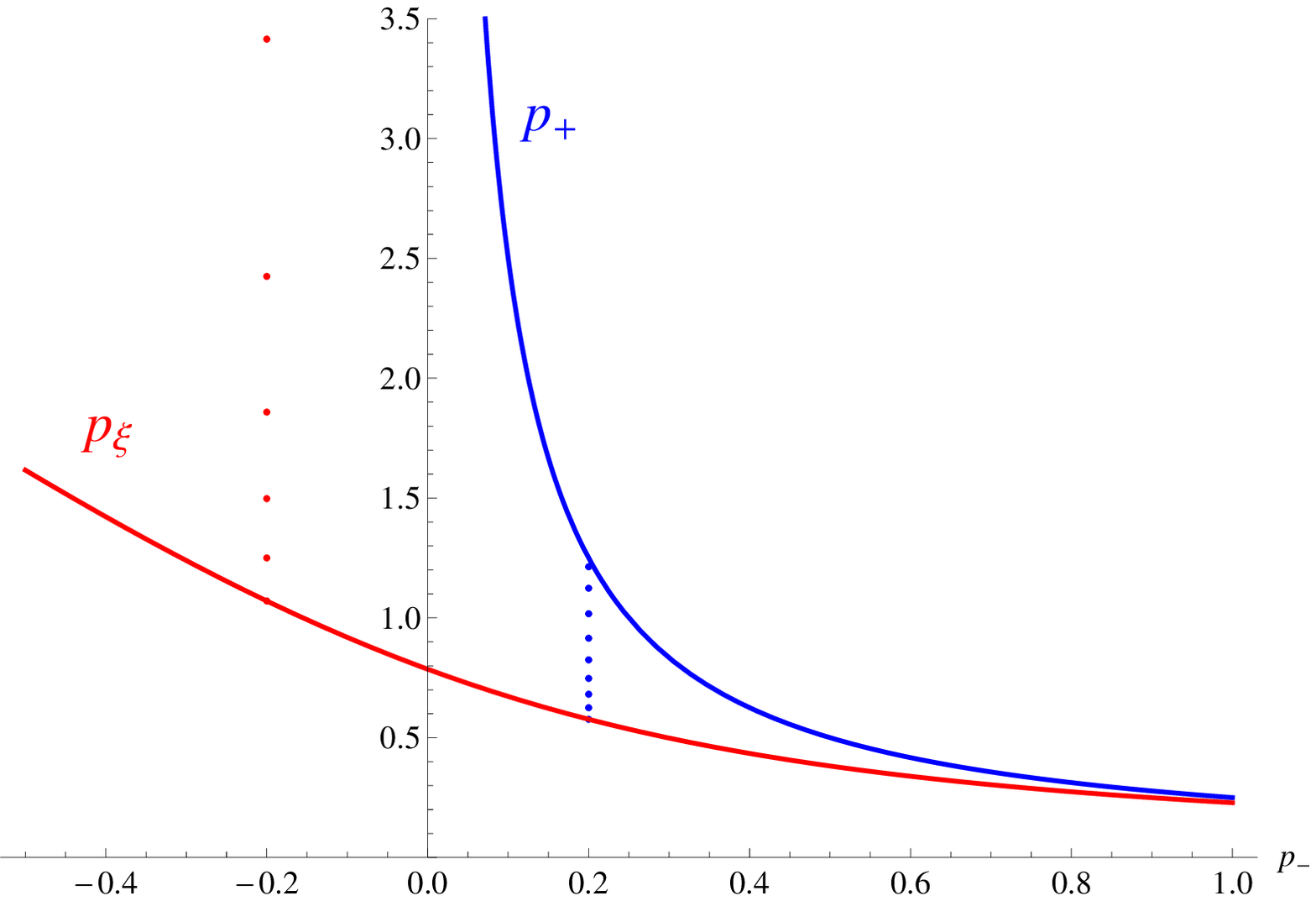}
	\caption{{\it Left}: Free particle energies $p_\xi$ in almost lightfront coordinates, as a function of $p_-$ (fixed $p_\perp$). $\eta$ decreases from red (dashed) to black (dotted) to blue (solid). Quadrants are labelled for later use.  {\it Right}: A state with energy $p_\xi$ in quadrant \ding{173} travels a finite spectral distance in the limit $\eta\to 0$, becoming a lightfront state with energy $p_+$. However, states in quadrant \ding{172} acquire an infinite energy and decouple from the theory.}
	\label{spec2}
\end{figure*}

We now turn to our second example of a limit in which spectral flow is at work. Consider the following (kinetic terms of) scalar field actions
\begin{eqnarray*}
	\mathcal{S}_2 &=& \frac{1}{2}\int\!\ud^4x\ \big\{ (\partial_0\phi)^2 -(\nabla\phi)^2 -m^2 \phi^2\big\}\;,\\
	\mathcal{S}_1 &=& \frac{1}{4}\int\!\ud x^+\ud^2 x^\perp \ud x^-\ \big\{4\partial_-\phi \partial_+\phi - (\nabla_\perp\phi)^2\\
	&&\hspace{4cm}-m^2\phi^2\big\}\;.	
\end{eqnarray*}
These are of course the same action, but the first is written in ordinary, instant form, coordinates $(x^0,x^j)$, while the second is written in lightfront coordinates $(x^+,x^\perp,x^-)$. We see that going from one to the other takes us from an action which is {\it quadratic} in time ($x^0$) derivatives to one which is {\it linear} in time ($x^+$) derivatives, a sign that there are connections between this system and that described above. In order to more closely examine what happens when we make the transition between these pictures, we will work in interpolating, `almost lightfront' coordinates \cite{Chen:1971yg}, which have found application in several areas \cite{Prokhvatilov:1989eq, Hellerman:1997yu, Ilgenfritz:2006ir}. To make life as simple as possible, we work with the usual lightfront spatial coordinates $x^-$ and $x^\perp$, but rather than taking $x^+$ as our time direction, we instead introduce a parameter $\eta$ and use the timelike coordinate $\xi$ defined by
\be\label{xi}
	\xi:= x^0\big(1+\frac{\eta^2}{2}\big) + x^3\big(1-\frac{\eta^2}{2}\big)\;.
\ee
As $\eta\to 0$, we rotate into lightfront coordinates proper, ie $\xi\to x^+$. Consider now the spectrum of a particle in our coordinates, found by extracting the on--shell energy $p_\xi$ (the momentum conjugate to $\xi$) from the mass--shell constraint $p_\mu p^\mu=m^2$,
\be\label{pxi}
	p_\xi = - \frac{p_-}{\eta^2}\ \pm \sqrt{\frac{p_-^2}{\eta^4} + \frac{p_\perp^2+m^2}{2\eta^2}}\;.
\ee
We see that $p_\xi$, unlike the lightfront energy $p_+$, contains a square root, and there are two solutions for each $p_\perp$ and $p_-$. The energies are plotted in Fig.~\ref{spec2}, left panel.

Now, we expect to recover the usual lightfront spectrum as we take the limit $\eta\to 0$. To see what happens, label the four quadrants of the $(p_-,p_\xi)$ plane as in Fig.~\ref{spec2}, left panel. In quadrants \ding{173} and \ding{175}, energies remain {\it finite} as $\eta\to 0$, and tend to lightfront energies,
\be
	p_\xi = \frac{p_\perp^2+m^2}{4p_-} + \mathcal{O}(\eta^2)\ {\longrightarrow p_+}.
\ee
In quadrants \ding{172} and \ding{174}, however, the energies blow up,
\be
	p_\xi = -\frac{2p_-}{\eta^2} - \frac{p_\perp^2+m^2}{4p_-} + \mathcal{O}(\eta^2)\ \longrightarrow \infty.
\ee
Half of the states in the theory therefore become infinitely excited and decouple in a spectral flow as $\eta\to 0$. The other half remain at finite energy, and become lightfront states, as illustrated in Fig.~\ref{spec2}, right panel.

What are the noncommutative consequences of this flow? Returning to $\eta\not=0$, the action is
\begin{eqnarray}\label{Leta}
	\nonumber \mathcal{S}_\eta =\frac{1}{4}\int\!\ud\xi\ud x^\perp\ud x^-\ \big\{2 \eta^2 \partial_\xi\phi\partial_\xi\phi + 4\partial_-\phi\partial_\xi\phi\\
	-(\partial_\perp\phi)^2-m^2\phi^2\big\}\;.
\end{eqnarray}
Quantising at $\eta\not=0$, we would infer the equal $\xi$ commutators
\begin{eqnarray}
	\nonumber [\phi(\xi,\boldsymbol{x}), \pi(\xi,\boldsymbol{y})] &=& i\delta^3(\boldsymbol{x}-\boldsymbol{y})\;, \\
	\qquad [\phi(\xi,\boldsymbol{x}), \phi(\xi,\boldsymbol{y})] &=& 0\;,
\end{eqnarray}
where $\x\equiv\{x^\perp,x^-\}$ and the momentum $\pi$ is easily read off from (\ref{Leta}). These commutators depend on cancellations between Fourier modes of the field, which in the quantum theory are particle creation/annihilation operators. Now, we know that half of the corresponding states decouple as $\eta\to 0$. The loss of these modes alters the commutators of the theory; there is now an incomplete cancellation between the modes and, as a result, the field--field commutator no longer vanishes at $\eta=0$. We obtain, of course, the lightfront field--field commutator \cite{Heinzl:2000ht}:
\begin{eqnarray}
	[\phi(\xi,\boldsymbol{x}), \phi(\xi,\boldsymbol{y})] \to [\phi(x^+,\x), \phi(x^+,\y)] \\
	\nonumber =-\frac{i}{4}\delta^2(x^\perp-y^\perp)\text{Sign}(x^- - y^-)\;.
\end{eqnarray}
So although the physics is very different, the same mechanism is at work here as was for the particle in a strong magnetic field. We have a limit in which high energy states decouple in a spectral flow. This results in configuration space being transformed into phase space. This is seen as the appearance of noncommutativity in configuration space (which here is field space, and for the previous example was $\mathbb{R}^2$).
We remark that the $\eta\to 0$ limit we have considered may be smooth or not, depending on what object is examined, and may have application to understanding zero modes and renormalisation in lightfront field theory -- for further discussion, see \cite{Heinzl:2007ca}.
\section{Other examples}
We very briefly mention two further limits which fit into our scheme. Spectral flow is of course at work in the decoupling limit of string theories with D--branes and magnetic backgrounds; closed strings and excited open string modes decouple, leaving a noncommutative field theory on the branes \cite{Chu:1998qz, Seiberg:1999vs, Ambjorn:2000yr}.

Another example is the non--relativistic limit of field theory, see e.g. \cite{zee:2003}, in which states with $E>m$ become inaccessible (the flow). In this limit the Klein--Gordon equation becomes the Schr\"odinger equation, so that for a scalar field the action goes from being quadratic in time derivatives to linear,
\bestar
	\int\!\ud^4x\ \phi^\dagger (-\partial^2-m^2)\phi \to \int\!\ud^4x\ \Phi^\dagger \bigg[i\partial_t+\frac{\nabla^2}{2m}\bigg]\Phi\;.
\eestar
(The change from $\phi$ to $\Phi$ is a simple phase shift needed in the limit.) Again, the result is noncommutativity in configuration (field) space: the commutator $[\Phi,\Phi^\dagger]$ is non--zero in the nonrelativistic limit. One can also include interactions and see that, because of the cutoff in energy--momentum space imposed by the non--relativistic condition, only those interactions which conserve particle number survive the limit \cite{Heinzl:2007ca}.

We now move on to our final example; the appearance of lightlike noncommutativity from a strong background field, which we will consider in a little more detail, and see that a significant role is played by lightfront zero modes.
\section{Lightlike noncommutativity}
We saw above that space--space noncommutativity appeared in the theory of a particle in a strong magnetic field. Are there other backgrounds which lead to other types of spacetime noncommutativity?  Consider a {\it crossed field}, which has orthogonal electric and magnetic fields of equal magnitude, i.e. $\E.\B=0$ and $|\E|=|\B|=F$, say. These are often employed as low frequency approximations to the electromagnetic fields of intense lasers. There is currently a great deal of interest in intense laser physics, as facilities such as Vulcan and ELI will allow us to probe nonlinear vacuum effects in QED \cite{Marklund:2008gj, Blaschke:2008wf, Heinzl:2008an}. The action for a particle in a crossed field may be written
\be\label{tofix}
	\mathcal{S}=\int\!\ud\tau\ -m\sqrt{\dot x^2} + e F x^+\,\dot{x}^1\;,\quad \dot{x}^\mu\equiv \frac{\ud x^\mu}{\ud\tau}\;,
\ee
where $\tau$ parameterises the particle's worldline and we have adopted a gauge with $A_1 = F x^+$ the only nonzero component of the potential. The relativistic particle action is needed for two reasons -- first, because in order to describe the crossed field we need to introduce a {\it lightlike} direction, here chosen to be $x^+$, and second, a particle in a crossed field is accelerated by the electric component, so that a relativistic description is eventually necessary.

If we now imagine that $e F\gg m^2$, so that we may neglect the kinetic term of the action, just as for the magnetic background, then we arrive at the non--zero commutator
\be\label{act}
		[x^+,x^1] = \frac{i}{eF}\;, 
\ee
between a lightlike and a spacelike direction, which goes by the name of lightlike noncommutativity \cite{Aharony:2000gz}. We recall that time--space noncommutative theories must be very carefully defined in order to avoid potential unitarity problems \cite{Gomis:2000zz}. However, for lightlike, as for space--space, noncommutativity, the usual Feynman diagram expansion yields a unitarity field theory \cite{Bahns:2002vm}. We also remark that scattering processes in laser backgrounds are specifically sensitive to lightlike noncommutativity, essentially because the laser photon momentum is a lightlike vector \cite{Heinzl:2009zd}.

We turn again to the energy spectrum. In order to construct a Hamiltonian for the relativistic particle, we need to gauge fix the reparametrisation invariance of (\ref{tofix}), which amounts to choosing a time direction. The method for doing so is well known, see \cite{Heinzl:2000ht}, and we will skip to the final result: we can gauge fix  $x^+\equiv\tau$, the worldline parameter, and then the Hamiltonian is $p_+$, which generates evolution in time $x^+$,
\be\label{H+}
		H_+ := p_+  = \frac{(p_1+ eF x^+)^2 + p_2^2 + m^2}{ 4p_{-}}\;.
\ee
This Hamiltonian is explicitly time dependent due to the term $eF x^+ \equiv eF \tau$, and reduces to the usual free particle lightfront Hamiltonian when $F=0$. Now, something strange has happened. We expected to find a theory with the commutator (\ref{act}), but from the outset we fixed $x^+\equiv\tau$, a smooth parameter: there can therefore be no such noncommutativity in a theory described by the Hamiltonian (\ref{H+}). To emphasise this point, let us look for the spectral flow which we should by now expect to be present if we are to see noncommutativity\footnote{We remark briefly that it is possible to consider this system in a completely time {\it independent} picture. We can gauge fix $x^-=\tau$ as time, in which case $x^+$ is spatial. This is perhaps a more natural starting point, as it disentangles the time and (expected) noncommuting directions from the outset, akin to the case of a magnetic field. However, the Hamiltonian in this picture is $p_-$, which has an ordering ambiguity and we have found it is actually more fruitful to adopt $x^+$ as time.}.
\subsection{Spectral flow and lightfront zero modes}
Quantising, $H_+$ commutes with both $p_\perp$ and $p_-$, so that the right hand side of (\ref{H+}) gives time--dependent `eigenvalues' $E_+$ of the Hamiltonian,
\be\label{E+}
	E_+(x^+) \equiv \frac{(p_1+ eF x^+)^2 + p_2^2 + m^2}{ 4p_{-}}\;,
\ee
which are equal to the time--dependent expectation values $\bra{p_\perp,p_-}H\ket{p_-,p_\perp}$. We see that as time $x^+$ increases, the energy of a particle increases like $(eFx^+)^2$. Physically, this is due to the electric field accelerating the charge, and this is also the source of the time dependence in the theory. If we impose a cutoff $\Lambda$ at some high energy scale, the energy of any particle will exceed this after a finite time.

Now, what happens as $F$ increases? We see that the larger $F$, the sooner a particle's energy reaches the cutoff scale $\Lambda$. Here we have our spectral flow; as $F$ increases,  all the modes described by (\ref{H+}) rapidly acquire energies above any given scale. As $F\to\infty$, {\it all} of the states described by (\ref{H+}) become infinitely excited, and there is nothing left behind to furnish us with a noncommutative theory. So we do have a flow, but seemingly no noncommutativity -- what is happening? The loophole in this argument is that there are modes in the theory {\it not} propagated by the Hamiltonian $H_+$: the notorious {\it zero modes} of $p_-$. If we are going to find any noncommutativity, we must find it amongst the zero modes. 

To pursue this, consider a complex scalar $\phi$ in our crossed field. The Klein--Gordon equation in this background can be solved exactly, as can the Dirac equation, giving the well known Volkov electrons \cite{volkov:1935}.  We note, though, that at the core of the Volkov solution is the adoption of lightfront coordinates, and as a result zero modes must be paid special attention. In particular, the propagator $\Delta$, which is a transition amplitude in the first quantised theory, has two `branches'; if $p_-$ is nonzero
\bestar\begin{split}
	\Delta_{p_-,p_\perp}(x^+,y^+) = &\frac{\theta\big(p_-(x^+-y^+)\big)}{4|p_-|}\\
	& \times \exp\bigg[ -i\int\limits_{y^+}^{x^+}\!\ud u^+ \ E_+(u^+)\bigg]\;,
\end{split}\eestar
whereas if $p_-=0$,
\bestar
	\Delta_{0,p_\perp}(x^+,y^+) =\frac{i\delta(x^+-y^+)}{p_2^2+(p_1+eF x^+)^2+m^2}\;.
\eestar
Note the similarity of this result to that of the free theory with $F=0$ \cite{Heinzl:2003jy}. In the above, $E_+$ is exactly as in (\ref{E+}), which shows us that the modes with $p_-\not=0$ are propagated by the particle Hamiltonian $H_+$, and we know from above that they see no noncommutativity. However, the zero mode propagator is very different -- it is instantaneous in $x^+$.

Do the zero modes see noncommutativity? When $m\not=0$, the zero modes can appear as internal lines of Feynman diagrams but they cannot go on shell -- the situation is just as for the free particle, where there is no way to satisfy the mass--shell condition if $p_-=0$ and $m\not=0$. Whether the resulting theory can be understood in terms of a noncommutative theory is an open question. (Related to this, see \cite{Miransky:2005zy} and references therein for a discussion of QED projected into the lowest Landau level.) However, when $m=0$, on--shell solutions do exist, which could have a quantum mechanical description; in the free theory, we take $p_\perp=p_-=0$ and then $p_+$ is arbitrary. In the interacting case, the Klein--Gordon equation is satisfied by taking $p_2 = p_1+eF x^+=0$, so that 
\be\label{phizero}
	\phi(x) = f(x^+) \exp\big( i e F x^1 x^+ \big)\;,
\ee
which is rather different to the Volkov scalars found at $p_-\not=0$.

Let us compare these results with those for a magnetic field. Assuming that all of the $p_-\not=0$ modes have decoupled, then all that remains are the zero modes. This is the analogue of the lowest Landau level projection which occurs in the magnetic case. In (\ref{phizero}), we have again killed the momentum component, here $p_2$, perpendicular to the plane which is expected to be noncommutative, here $(x^+,x^1)$. The modes are then defined by $p_1=-eF x^+$ which indeed gives us an effective noncommutative description respecting (\ref{act}). There is again an infinite degeneracy of states, seen here in the arbitrary function $f(x^+)$. We clearly have some similarities with the case of a strong magnetic field, although the physics is quite different due to the time--dependence of the theory, and the role played by lightfront zero modes. We hope to expand on this discussion in a future publication.
\section{Conclusions}
We have seen that many limits can be unified by spectral flow, in which a decoupling of states transforms configuration space into phase space and so introduces a noncommutativity into the theory. We reviewed the examples of a particle in a strong magnetic field, quantising in almost lightfront coordinates, strings in magnetic backgrounds and the non--relativistic limit of field theory.

Finally, we considered the appearance of lightlike noncommutativity from a strong crossed field. We found some similarities with the case of a strong magnetic field, but also saw that there is an intimate connection between this limit and lightfront zero modes. In particular, it seems that if a noncommutative theory emerges in the strong crossed field limit, it can be a theory {\it only} of the zero modes.

\subsubsection*{Acknowledgements}
A.~I. thanks the organisers of LightCone 2009 for the opportunity to attend, the participants for many interesting discussions, and Richard Szabo for very useful correspondence.

A.~I. is supported by IRCSET. M.~M. is supported by the European Research Council under Contract No.\ 204059-QPQV, and by the Swedish Research Council under Contract No.\ 2007-4422.

\end{document}